\documentclass[aps,prl,twocolumn,longbibliography,10pt]{revtex4-1}

\usepackage{amsmath}
\usepackage{amssymb}
\usepackage{graphicx}
\usepackage{fancyhdr}
\usepackage{lastpage}
\usepackage{color}
\usepackage[normalem]{ulem}
\usepackage[utf8]{inputenc}
\usepackage[T1]{fontenc}
\usepackage{lmodern}
\usepackage{bm}  
\usepackage{bbold}

\begin{document}

\title{Spontaneous formation and relaxation of spin domains in antiferromagnetic spin-1 quasi-condensates}

\author{K. Jim{\'e}nez-Garc{\'i}a, A. Invernizzi, B. Evrard, C. Frapolli, J. Dalibard and F. Gerbier}
\affiliation{Laboratoire Kastler Brossel, Coll{\`e}ge de France, CNRS, ENS-PSL Research University, Sorbonne Universit{\'e}, 11 Place Marcelin Berthelot, 75005 Paris, France}

\date{\today}

\maketitle


\section{Abstract}
\textbf{Quantum systems of many interacting particles at low temperatures generally organize themselves into ordered phases of matter, whose nature and symmetries are captured by an order parameter. In the simplest cases, this order parameter is spatially uniform. For example, a system of localized spins with ferromagnetic interactions align themselves to a common direction and build up a macroscopic magnetization on large distances. However, non-uniform situations also exist in nature, for instance in antiferromagnetism where the magnetization alternates in space. The situation becomes even richer when the spin-carrying particles are mobile, for instance in the so-called stripe phases emerging for itinerant electrons in strongly-correlated materials. Understanding such inhomogeneously ordered states is of central importance in many-body physics.
In this work, we study experimentally the magnetic ordering of itinerant spin-1 bosons in inhomegeneous spin domains at nano-Kelvin temperatures. We demonstrate that spin domains form spontaneously after a phase separation transition, \textit{i.e.} in the absence of external magnetic force, purely because of the antiferromagnetic interactions between the atoms. Furthermore, we explore how the equilibrium domain configuration emerges from an initial state prepared far-from-equilibrium.
}

\section{Introduction}

Quantum gases of ultracold atoms offer an unprecedented platform to study complex, multi-component quantum fluids in- and out-of-equilibrium~\cite{pitaevskiibook,kawaguchi2012a,stamperkurn2013a}. An example is provided by bosonic quantum gases with several Zeeman components simultaneously confined in an optical dipole trap, where Van der Waals\,\cite{chang2004a,Schmaljohann2004,kuwamoto2004,chang2005,kronjager2006,black2007a,pechkis2013} or dipole-dipole\,\cite{lahaye2009a,pasquiou2011a} interactions drives internal conversion between the Zeeman components. For bosonic atoms, this leads at very low temperatures to Bose-Einstein condensation in a superposition of the internal states (a so-called \textit{spinor condensate}) where long-range phase coherence, superfluidity and magnetic ordering can all take place. For instance, Josephson-like spin oscillations due to spin-changing interactions have been observed experimentally~ \cite{chang2005,kronjager2006,black2007a}, and spin superfluidity demonstrated in recent experiments with sodium atoms~\cite{fava2017a,Kim2017a}.

A major question that arises for multi-component fluids --quantum or classical-- is the stability of spatially homogeneous phases towards phase separation~\cite{pitaevskiibook}. In cold atom experiments, phase separation has been observed in numerous multicomponent systems, either in dual species Bose-Bose or Bose-Fermi mixtures \cite{modugno2002a,guenter2006a,ospelkaus2006a,ferlaino2006a,Thalhammer2008,Papp2008, McCarron2011,Wacker2015,desalvo2017a} or for single species quantum gases with several hyperfine components, \textit{e.g.} two-component imbalanced Fermi gases with strong interactions~\cite{shin2006a} or bosonic mixtures of hyperfine states \cite{Myatt1997,Hall1998,De2014,nicklas2015a}. Reaching equilibrium in dual species mixture can be difficult if inelastic losses are strong, \textit{e.g.} near a Feshbach resonance. In that context, metastable phase-separated configurations were reported in~\cite{Papp2008}. Furthermore, in many cases the different components experience different trapping potentials due to different magnetic moments or masses. A species- or spin-dependent trapping potential can strongly influence phase separation in a trapped system, to the point where it becomes the main factor deciding its occurrence instead of interatomic interactions~\cite{pitaevskiibook,lee2016a}.

\begin{figure*}[hhhhhttttt!!!!!]
\includegraphics[width=\textwidth]{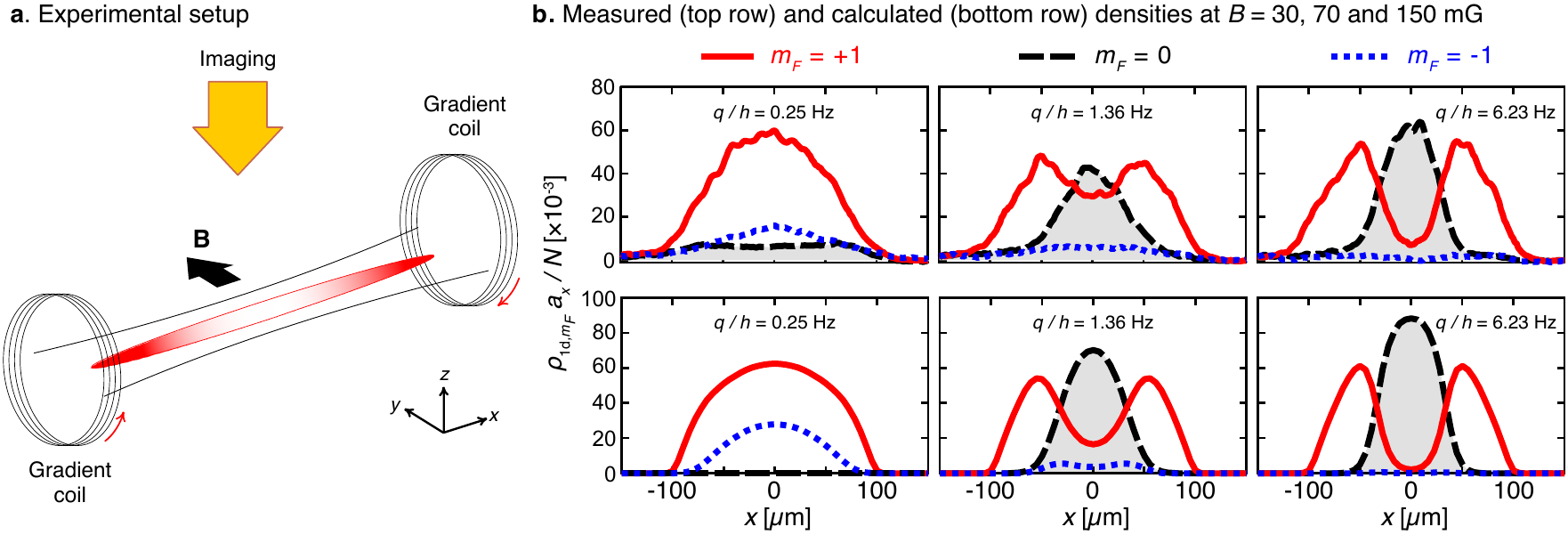}%
\caption{{\bf Spin domain formation in a quasi-1d spinor gas without applied magnetic force.} {\bf a.} Sketch of the experiment. A quasi-one-dimensional Bose-Einstein condensate of spin-1 Sodium atoms is immersed in a spatially uniform magnetic field $\bm{B}$. We use the compensation coils to cancel stray magnetic field gradients along the long axis of the cloud, thereby suppressing external magnetic forces. {\bf b.} Linear density profiles for increasing magnetic field. The top row shows the experimental profiles, obtained by averaging over about 100, 40 and 30 individual profiles, respectively. For low applied magnetic fields, we find that the Zeeman components $m_F=\pm 1$ coexist, with $m_F=0$ atoms forming a broad, presumably thermal background. When the magnetic field increases, we observe the formation and growth of a $m_F=0$ domain at the center of the trap. The bottom row shows theoretical profiles at $T=0$, calculated by solving numerically the one-dimensional spin-1 Gross Pitaevskii equation (see Methods), in good agreement with the observed profiles. The average profiles are symmetric under reflection, as expected in the absence of magnetic gradients along $x$. The trap frequencies are $(\omega_x,\omega_\perp)=2\pi\times(3.1,270)\,$Hz, the longitudinal magnetization is $M_{\|}\approx 0.5\,N$ and the atom number is $N\approx 10^4$.}
\label{fig:Domains}
\end{figure*}

In this work, we study the formation of spin domains in a quasi-one-dimensional (1d) spinor Bose-Einstein condensate (BEC) in an external, spatially uniform magnetic field without any external magnetic force. The condensate is made from sodium atoms carrying an hyperfine spin $F=1$. The spin-dependent interactions have an antiferromagnetic character that leads to phase separation~\cite{stenger1998a,isoshima1999a,miesner1999a,stamperkurn1999b,kronjager2010a}. Early experiments observed spin domains in a $F=1$ sodium BEC immersed in a magnetic field gradients around $10\,$mG/cm~\cite{stenger1998a,miesner1999a,stamperkurn1999b}. 
The magnetic force produced by the gradient make the $m_F=\pm 1$ Zeeman components migrate to opposite sides of the trap, with the $m_F=0$ component in-between. Without applied gradient, only the miscible $m_F=\pm 1$ phase was observed in~\cite{stenger1998a}. 

Refs.~\cite{matuszewski2008a,matuszewski2009a,matuszewski2010a} pointed out theoretically that $m_F=0$ spin domains should also form without any applied gradient. For a gas in a box, the domain should preferentially move to one side of the box to have only one interface, with $m_F=0$ on one side of the box and $m_F=\pm 1$ on the other. For a trapped gas, the energetic cost of the ``additional'' interface is compensated by the gain in the interaction energy when the $m_F=0$ domain is located in the center of the trap (see~\cite{matuszewski2010a} and below). 

The spin-1 quantum gas in our experiments is confined in a spin-independent and highly elongated trap, realizing an effectively 1d spinor gas where phase separation occurs only along the weak axis. We take special care to compensate magnetic field gradients along that axis (cancelling them below the mG/cm level) to ensure the domains form in a negligible magnetic force. We measure the equilibrium spatial distributions, which reflect (up to interface effects that we quantify) the phase boundaries for systems with homogeneous particle density. We find qualitative agreement but quantitative differences between the measured equilibrium distributions and $T=0$ mean-field theory. We attribute these differences to thermal fluctuations, which play an important role due to the low-energy scales associated with spin ordering, and the low dimensionality.

Another important question besides the nature of the equilibrium state is whether this state can be reached on a timescale compatible with the lifetime of the atomic sample. Refs.\,~\cite{miesner1999a,stamperkurn1999b} studied relaxation in a strong applied magnetic field gradient, observing that metastable configurations can persist for seconds. Several experiments, mostly using $F=1$ $^{87}$Rb atoms with ferromagnetic interactions~\cite{sadler2006a,MurPetit2006a,saito2007a,MurPetit2009a,kronjager2010a,Vengalattore2010a,De2014,nicklas2015a}, studied the dynamical formation of non-equilibrium spin domains after a quench. Reaching an equilibrium state appears difficult for $^{87}$Rb atoms due to the weakness of spin interactions~\cite{guzman2011a}. Other experiments with $F=1$ sodium atoms observed the formation of short-lived domains across a quantum phase transition and studied their equilibration dynamics~\cite{bookjans2011a,kang2017a}. However, heating due to the experimental arrangement prevented to study the long-time regime and the approach to the expected equilibrium state. The relation between the formation of spin domains after a quench and the Kibble-Zurek mechanism has also been discussed~\cite{swislocki2013a}. In the second part of the article, we adress the issue of relaxation to equilibrium in a gradient-free situation. We prepare a spin configuration far from equilibrium and monitor how it relaxes to equilibrium. We observe a slow relaxation on a time scale of several seconds, and a spin dynamics that points to spin-mixing collisions as the underlying relaxation mechanism. 

\section{Results}

\begin{figure*}[ht!!!!]
\includegraphics[width=\textwidth]{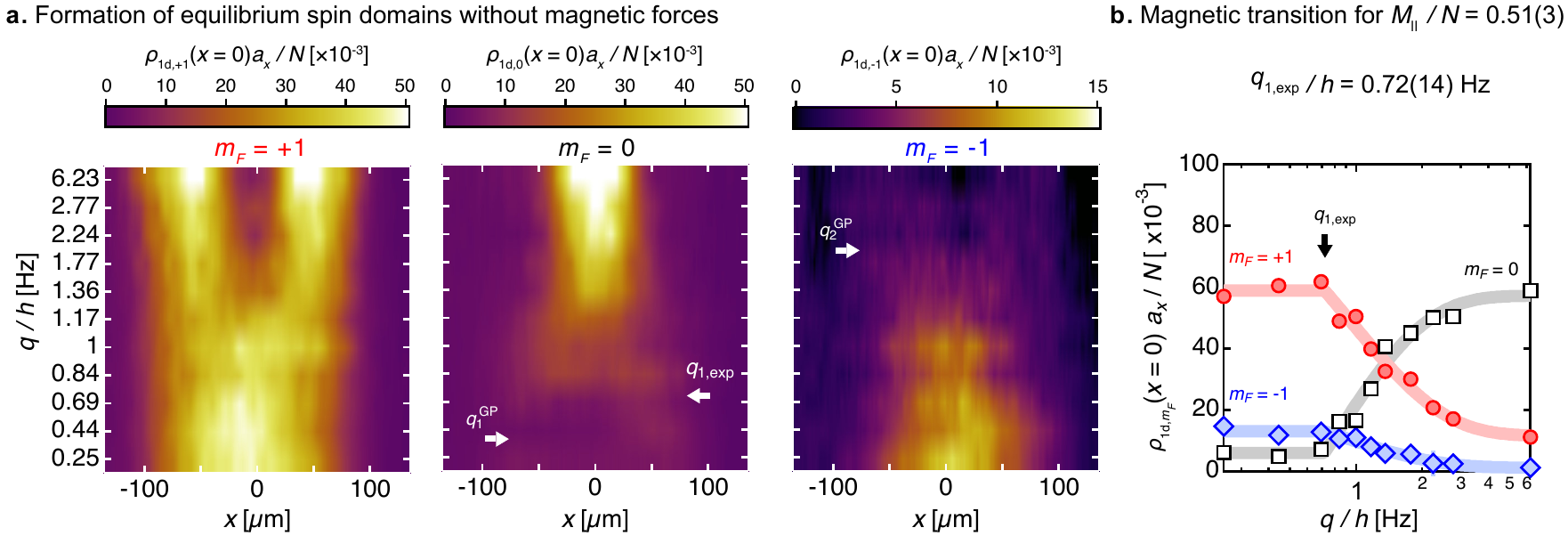}%
\caption{{\bf Magnetic phase transition and phase boundaries.} {\bf a.} Linear densities of each Zeeman component $m_F$ as a function of position for increasing quadratic Zeeman energies $q$. Each line in the false color plots corresponds to an individual profile as shown in Fig.\,\ref{fig:Domains}b. The arrows mark the locations of the observed $q_{1,\textrm{exp}}$ and predicted $q_1^{\textrm{GP}},q_2^{\textrm{GP}}$ critical quadratic Zeeman energies. {\bf b.} Evolution with $q$ of the normalized densities at the center of the trap. The gray lines show a piece-wise function constant below $q_{1,\textrm{exp}}$ and growing as $1-e^{-\vert q-q_{1,\textrm{exp}}\vert/\Delta q}$ above. We obtain the quoted experimental values of $q_{1,\textrm{exp}}$ fitting this function to the experimental data for $m_F=0$. The quoted uncertainty correspond to a 90\,\% confidence interval assuming Gaussian noise and independent errors. The trap frequencies, longitudinal magnetization and atom number are as in Fig.\,\ref{fig:Domains}.}
\label{fig:PhaseTransition}
\end{figure*}

\subsection{Experimental System}

Our experiments are performed with a gas of $^{23}$Na atoms trapped in a spin-independent optical trap with frequencies $\omega_x,\omega_\perp$ along the weak and strong axes, respectively (Fig.~\ref{fig:Domains}\textbf{a}). 
With a total atom number around $N \approx 10^4$, the chemical potential of a single component gas at low temperatures is $\mu \sim 0.5 \hbar\omega_\perp$. This implies a quasi-one dimensional (1d) regime where transverse motion is almost frozen to the ground state of the harmonic potential. The measured $1/e$ lifetime of the cloud is around $50\,$s, presumably limited by residual evaporation and three-body recombination.

We measure the linear integrated densities $\rho_{1{\rm d},m_F}(x)=\int dydz\,\rho_{m_F}({\bf r})$ along the weak axis of the trap after a short expansion in a magnetic field gradient that separates all three components $m_F=0,\pm 1$ by the Stern-Gerlach effect. Here and in the following, $\rho_{m_F}$ denotes the partial density of the Zeeman component with magnetic quantum number $m_F=0,\pm 1$, $\rho = \sum_{m_F} \rho_{m_F}$ the total density, and the subscript ``1d'' always indicates linear quantities integrated over the transverse coordinates $y,z$. 

The quasi-1d character of the trapped gas results in spatial fluctuations of the phase of the order parameter along the weak axis of the trap~\cite{petrov2001a}. Such fluctuations do not affect significantly the thermodynamic properties of the mixture, but they show up as density stripes in time-of-flight images~\cite{dettmer2001a}. The density profiles reported in this article are averaged profiles over many (typically several tens) repetitions of the experiment to suppress the signature of phase fluctuations. Because of the very weak expansion along $x$, the observed average distributions reflect the linear in-trap density distributions to a good approximation. We also take special care to cancel residual magnetic forces that could affect the spatial distributions (see Methods). This is reflected in the nearly symmetric linear distributions of the spin components (Fig.~\ref{fig:Domains}\textbf{b}).

\subsection{Brief Review of Ultracold Spin-1 Gases}

Before discussing our results, we first review the salient features of $F=1$ spinor condensates \cite{stamperkurn2013a}. At very low temperatures, Bose-Einstein condensation leads to a macroscopic occupation of a single-particle state ${\bf \Psi}$, a superposition of all three Zeeman states behaving as a three-dimensional vector. 
The equilibrium many-body state is determined by the competition between the interatomic interactions and the Zeeman energy in an applied magnetic field. 
The total mean-field energy at $T=0$ takes the form \cite{stamperkurn2013a}
\begin{align}
E &=\int d{\bf r}\, \left( {\bf \Psi}^\ast \hat{h} {\bf \Psi}+\frac{\overline{g}}{2} \rho^2+\frac{g_s}{2} {\bf m}^2 \right).
\end{align}
Here $\hat{h}=-\frac{\hbar^2}{2m_{\textrm{Na}}}\Delta\cdot +V+E_{\textrm{Zeeman}}$ is the single-particle Hamiltonian, $m_{\textrm{Na}}$ is the atomic mass, $E_{\textrm{Zeeman}}$ is the Zeeman energy discussed below, and $V=\frac{1}{2}m_{\textrm{Na}}  [\omega_x^2 x^2 + \omega_\perp^2 (y^2+z^2)]$ the trapping potential. The partial densities are given by $\rho_{m_F}=\vert \Psi_{m_F} \vert^2$. The magnetization density ${\bf m}$ is defined by its Cartesian components $m_\alpha = \sum_{i,j} \Psi_i^\ast ( \hat{S}_\alpha )_{i,j} \Psi_j$, with $\hat{S}_\alpha$ ($\alpha=x,y,z$) the standard spin-1 matrices \cite{stamperkurn2013a}. 

The two coupling constants $\overline{g}$ and $g_s$ characterize spin-independent and spin-dependent interactions, respectively. For sodium atoms in the $F=1$ hyperfine manifold the spin-dependent interactions are antiferromagnetic ($g_s>0$), a key feature to observe phase separation \cite{stenger1998a}. Furthermore, the spin-dependent term $\propto g_s$, although much weaker than the dominant spin-independent term ($ g_s / \overline{g} \approx 0.036$), is essential to understand spinor gases\,:\, This term lifts spin degeneracies left by $\overline{g}$ and determine the magnetic properties at very low temperatures.

Spinor gases are typically immersed in a uniform magnetic field ${\bf B}$ that shifts the internal energy levels by the Zeeman effect. The interaction Hamiltonian conserves the total longitudinal magnetization $M_{\|}=\int d{\bf r}\,m_{\|}$ with $m_{\|}$ the component of ${\bf m}$ along the axis of the applied magnetic field ${\bf B}$. As a result, the constant of motion $M_{\|}$ should be viewed as an experimental control parameter and not as a dynamical variable. The conservation of $M_{\|}$ makes the first-order Zeeman shift linear in $B$ irrelevant to the equilibrium properties. The relevant shift comes from the second-order or quadratic Zeeman energy, $E_{\rm Zeeman} = -q \int d{\bf r}\, \rho_0$ (up to a constant), with $q= \alpha_q {\bf B}^2$ and $\alpha_q \approx h\times 277\,$Hz/G$^2$ for sodium atoms.

\subsection{Magnetic Phase Diagram and Spontaneous Phase Separation}

We explore in Fig.\,\ref{fig:PhaseTransition} the equilibrium spatial structure of a quasi-1d antiferromagnetic spin-1 Bose gas in a spatially uniform applied field $\bm{B}$. We set the total longitudinal magnetization to $M_{\|}\approx 0.5\,N$ and vary the quadratic Zeeman energy (QZE) $q$. We find that the spatial structure of the spinor gas undergoes a marked change as $q$ increases. For low $q$, we observe a mixed phase where $m_F=\pm 1$ coexist in the same region of space in the center of the trap, surrounded by magnetized $m_F=+1$ regions near the edges of the cloud. Above a critical QZE $q_{1,\textrm{exp}}\approx h \times 0.72(14)~$Hz (corresponding to a magnetic field $B_{1,\textrm{exp}} \approx 51(5)~$mG), the $m_F=0$ component appears and develops into a domain expelling $m_F=\pm 1$ from the central region. The quoted experimental value of $q_{1,\textrm{exp}}$ is found by fitting an empirical function \----constant below $q_{1,\textrm{exp}}$ and growing as $1-e^{-\vert q-q_{1,\textrm{exp}}\vert/\Delta q}$ above\---- to the $m_F=0$ density in the trap center (Fig.\,\ref{fig:PhaseTransition}b). Error bars denote the $90\,$\% uncertainty level of the fit obtained by standard error analysis assuming Gaussian noise. Furthermore, for $q \gtrsim h \times 2~$Hz ($B \gtrsim 85\,$mG), the mixed $m_F=\pm 1$ region essentially disappears and the spin-1 gas reduces to a binary mixture of $m_F=0,+1$. Our data are summarized in Fig.\,\ref{fig:PhaseTransition}\textbf{a}, where we plot the linear partial densities $\rho_{1d,m_F}$ versus $q$. A similar behavior is observed for other values of the longitudinal magnetization $M_{\|}$. 

Besides the stripes due to phase fluctuations discussed earlier, we also observe substantial position fluctuations of the spin domains. For instance, in the examples shown in Fig.~\ref{fig:Domains}\textbf{b}, we find that the center-of-mass of the $m_F=0$ component fluctuates by $\sim 16\,\mu$m for $B=45\,$mG and by $\sim 6\,\mu$m for $B=150\,$mG. We believe this behavior is due to thermal fluctuations of the domain, and not to a technical artifact such as a magnetic gradient fluctuating around the compensated value. The fluctuations of the position of the spin domains and their possible use for low-temperature thermometry will be explored in more detail in a future publication.

\subsection{Phase Coexistence in Homogeneous Systems}

\begin{figure*}[t]
    \centering
\includegraphics[width=0.9\textwidth]{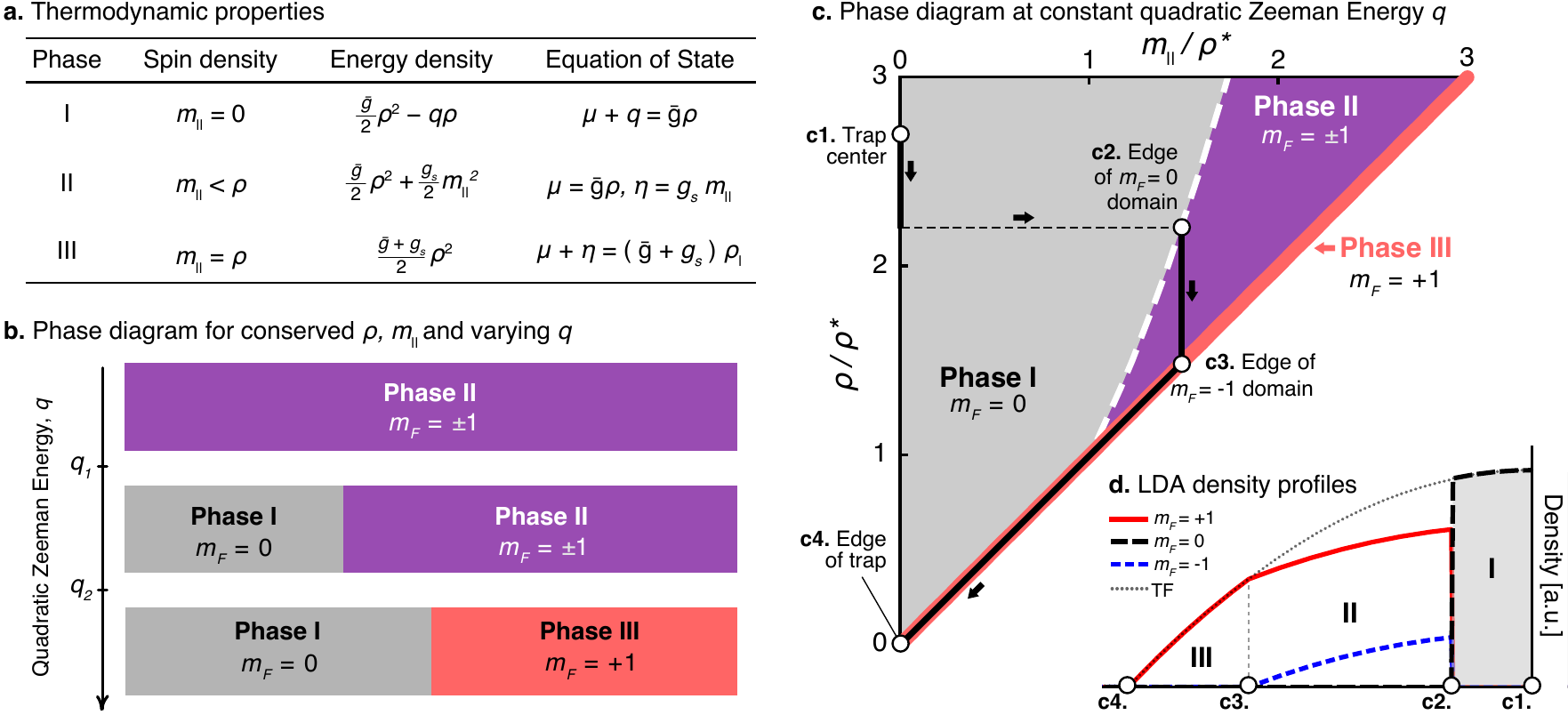}
\caption{\textbf{Phase diagram of homogeneous spin-1 systems for a fixed $q$.} {\bf a.} Summary of the thermodynamic properties of the posssible homogeneous phases of a spin-1 antiferromagnetic BEC. {\bf b.} Phase diagram for fixed density $\rho$ and longitudinal magnetization $m_{\|}$ as a function of the quadratic Zeeman energy $q$. {\bf c.} Phase diagram for fixed $q$ as a function of $\rho, m_{\|}$. The density scale is $\rho^\ast=2q/g_s$. The black line shows a typical ``trajectory'' from the trap center to the cloud edge for a trapped system treated in the local density approximation. {\bf d.} Density profiles of the Zeeman components in a trap in the local density approximation, corresponding to the trajectory shown in {\bf c}. The dotted line shows the total density.
}
\label{fig:homog}
\end{figure*}

The observed characteristics of the phase diagram can be qualitatively understood by considering first a uniform system in the thermodynamic limit enclosed in a box of volume $\mathcal{V}$. 
Three homogeneous phases can be realized depending on the magnetization $M_{\|}=m_{\|}\mathcal{V}$ \,\cite{stenger1998a,isoshima1999a,matuszewski2008a,matuszewski2010a},
\begin{itemize}
\item[] \textbf{Phase I {\rm or} Unmagnetized phase} --  All atoms occupy the $m_F=0$ Zeeman state, with $m_{\|}=0$ and $\rho_0 =\rho$,
\item[] \textbf{Phase II {\rm or} Partially magnetized phase} -- The components $m_F=\pm 1$ coexist, with magnetization density $0 < m_{\|} < \rho$ and $\rho_0=0$,
\item[] \textbf{Phase III {\rm or} Fully magnetized phase} -- All atoms occupy the $m_F=+1$ Zeeman state, with $m_{\|}=\rho$ and $\rho_0 =0$. Note that phase II evolves continuously into phase III when the magnetization increases. 
\end{itemize}
The properties of the various phases are summarized in Figure\,\ref{fig:homog}\textbf{a}. A completely homogeneous phase where the three Zeeman components coexist is always unstable towards phase separation \cite{matuszewski2008a}. For a partially magnetized system with $0 < M_{\|} < N$, phase II is the only possible homogeneous phase compatible with the conservation of the total magnetization $M_{\|}$. However, it competes with inhomogeneous (phase-separated) configurations, either ${\rm I}-{\rm II}$ or ${\rm I}-{\rm III}$, depending on the value of $q$ \cite{matuszewski2008a}. 

A common choice in the literature (made, \textit{e.g.} in Refs.~\cite{stenger1998a,matuszewski2008a}) is to describe the evolution of the system for fixed $\rho,~m_{\|}$ and with varying QZE $q$. For low QZE, phase II minimizes the interaction energy and is the stable equilibrium phase. As the QZE increases, a mixed configuration where part of the system is in phase II and part in phase I becomes energetically competitive. The preferred equilibrium configuration can be determined by comparing the energies of the competing possibilities (neglecting the energy cost of the ${\rm I}-{\rm II}$ interface),
\begin{align}
\delta E & = E_{{\rm I}-{\rm II}} -E_{\rm II} = \mathcal{V} f_0 \times\left[ \frac{g_s m_{\|}^2}{2 (1-f_0)} - q \rho\right],
\end{align} 
with $f_0$ the fraction of the available volume occupied by phase I in the mixed configuration. When $q \geq q_1=g_s m_{\|}^2/(2\rho)$, $\delta E$ becomes negative for $f_0=0$ and the homogeneous phase II becomes thermodynamically unstable. Above $q_1$, a phase I domain forms. The equilibrium fraction of $m_F=0$ atoms grows as $f_0 (q\geq q_1) =  1-(q_1/q)^{1/2}$. The conservation of the total magnetization $M_{\|}$ then requires that the magnetization density in sub-region II decreases as $m_{\|} = M_{\|}/[\mathcal{V}(1-f_0)]$. When $m_{\|}=\rho$ ($f_0 =  1-M_{\|}/N$), one obtains a phase-separated ${\rm I}-{\rm III}$ mixture which remains the same when $\rho$ increases further. 
The sequence of transitions is illustrated in Fig.\,\ref{fig:homog}{\textbf b}.

Anticipating the discussion of the trapped case within the framework of the local density approximation, we now adopt a slightly different point of view and consider the properties of the system for fixed $q$ and varying $\rho,~m_{\|}$ (Fig.\,\ref{fig:homog}\textbf{c}). It is convenient to chose a thermodynamic ensemble characterized by a chemical potential $\mu$ and a ``thermomagnetic'' potential $\eta$ conjugate to $N$ and $M_{\|}$, respectively. The equation of state of the various phases are given in terms of $\mu$ and $\eta$ in Figure\,\ref{fig:homog}\textbf{a}. Phase II (respectively phase I) is the stable equilibrium phase for densities below (resp. above) a critical value defined by
\begin{align}\label{eq:rhoq1}
\left( \rho q \right)_1 =  \frac{\eta^2}{2 g_s}=\frac{g_s m_{\|}^2}{2}
\end{align}
where $\eta=g_s m_{\|}$ in phase II. A second, continuous II-III transition occurs at $m_{\|} = \rho^\ast$, with the characteristic density
\begin{align}\label{eq:rho2}
\rho^\ast = \frac{2q}{g_s},
\end{align}
with the fully magnetized phase III realized for densities lower than $\rho^\ast$. 

\subsection{Spatial Structure of a Trapped System}

\begin{figure}[ht!!!!!]
\includegraphics[width=0.5\textwidth]{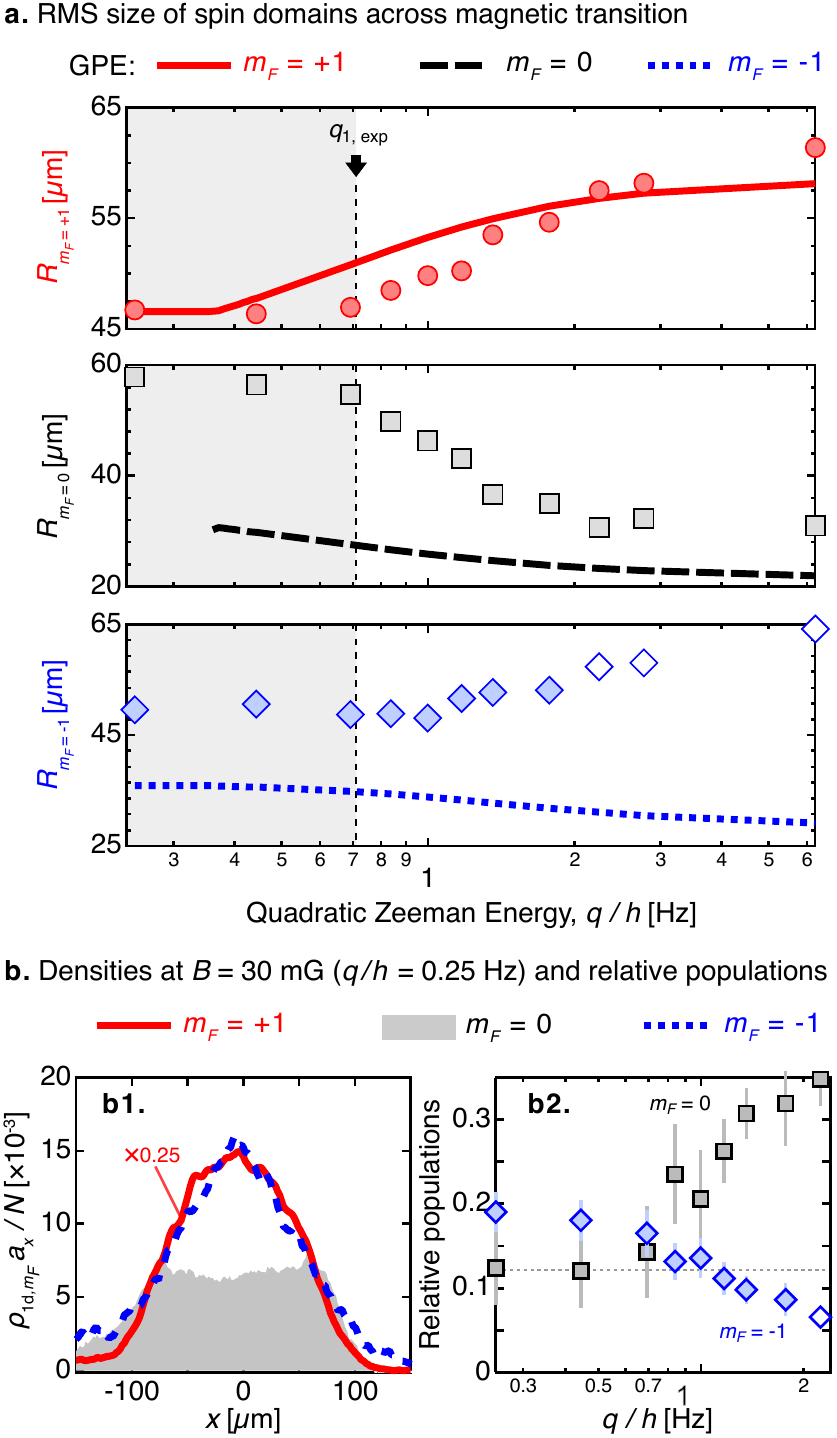}%
\caption{{\bf Domain size and role of the thermal component.} {\bf a.} Measured (symbols) and calculated (lines) effective size for each Zeeman component versus $q$. Open symbols indicate situations where the atomic density in $m_F=-1$ becomes comparable to the imaging background noise (total population below $\sim\,700$ atoms). We ascribe the differences between the measured and calculated $R_{M_F=0,-1}$ to the presence of a substantial thermal component. {\bf b.} Density profiles of the $m_F=0$ (gray), $m_F=+1$ (red) and $m_F=-1$ components (blue) below $q_{1,\textrm{exp}}$. The $m_F=0$ component exhibits a flat-top profile within the condensate region. The profile is the same as in Fig.\,\ref{fig:Domains}b, $q/h=0.25\,$Hz, with the $m_F=+1$ component scaled down by a factor $0.25$ for clarity. {\bf c.} Total populations of the $m_F=0$ and $m_F=-1$ components. Error bars indicate the empirical standard deviations of the data calculated over several tens of profiles for each $q$. The trap frequencies, longitudinal magnetization and atom number are as in Fig.\,\ref{fig:Domains}.}
\label{fig:Fits}
\end{figure}
The preceding discussion is directly relevant to determine the spatial structure of a quasi-1d gas in a harmonic trap where $\rho$, $m_{\|}$ and $\rho_0$ vary with position. We consider in this Section the purely 1d limit $\mu \ll \hbar\omega_\perp$ where the transverse motion is frozen in the transverse ground state of the trap. We first  perform our analysis within the local density approximation (LDA), and discuss effects beyond the LDA at the end of this section. The equalities established in the previous section remain valid substituting linear densities $\rho \to \rho_{1{\rm d}}$ and $(\overline{g},g_s) \to (\overline{g}^{\textrm{1d}},g_s^{\textrm{1d}})$, with effective 1d coupling constants $(\overline{g}^{\textrm{1d}},g_s^{\textrm{1d}})=(\overline{g},g_s)\times 1/(2\pi a_\perp^2)$. Here $a_\perp=\sqrt{\hbar/(m_{\textrm{Na}}\omega_\perp)}$ is the transverse harmonic oscillator size. Because the magnetization density $m_{\|}$ depends only on $x$, we keep the same notation  $m_{\|}$ for its integrated version with a slight abuse of notation. The pure 1d limit is not strictly realized in our experiment, as noted earlier. However, we have evaluated corrections to this limit and found that they only change marginally the conclusions (see Supplementary Material). As a result we stick to the 1d description in the core of the article to keep the discussion as simple as possible.


For given $N,M_{\|}$ the condition for the appearance of phase I in the center of the trap given in Eq.~(\ref{eq:rhoq1}) can only be fulfilled for sufficiently high QZE $q$. Similarly to the homogeneous case, this leads to a first critical value $q_1$ that corresponds to our measured $q_{1,\textrm{exp}}$. The magnetization density $m_{\|}$ in region II is uniform, but not directly proportional to $M_{\|}$ as it was in the homogeneous case. For a purely 1d system, we find following Ref.\,\cite{swislocki2013a} that $m_{\|}\approx \rho_{1{\rm d}}(0)[1-(1-(M_{\|}/N)^{2/3}]^2$ for $q \leq q_1$. Using Eq.~(\ref{eq:rhoq1}), this gives the LDA prediction for the first critical QZE~\cite{swislocki2013a},
\begin{align}\label{eq:q1LDA}
q_{1,\textrm{LDA}} =  \frac{g_s \mu}{2 \overline{g}}\left[1-\left(1-\frac{M_{\|}}{N}\right)^{2/3}\right]^2. 
\end{align}
Using our experimental parameters ($\mu/h \approx 120\,$Hz), we obtain $q_{1,\textrm{LDA}} \approx h\times 0.3\,$Hz, substantially below the observed $q_{1,\textrm{exp}}\approx h \times 0.72~$Hz. The same conclusion holds when taking the deviations from the purely 1d case into account (see Supplementary Material). 

The quantitative difference between the observations and the LDA prediction can be expected, as the latter completely neglects the energy cost of the domain wall between two immmiscible phases. This cost comes from the balance between the kinetic energy, increasing when the domain wall becomes steep, and the interaction energy, increasing when the wall spreads out due to the increased overlap between the two components. The energy of the domain wall is proportional to its width (typically several times the \textit{spin healing length} $\zeta_s = \sqrt{\hbar^2 \overline{g}/(2m_{\textrm{Na}}g_s \mu)} \sim7\,\mu$m), and therefore not extensive and negligible for infinitely large systems. However it can be significant in a gas of finite extent as in our experiment where a typical cloud half-length is $L=\sqrt{2 \mu/(m\omega_x^2)} \sim 100\,\mu$m. 

These effects beyond the LDA can be explored at zero temperature using the mean-field theory of spin-1 gases, which takes the form of three coupled Gross-Pitaevskii equations. We have solved these equations numerically to find the lowest energy solution (see Methods). Examples of the density profiles that we obtain numerically are shown in Fig.\,\ref{fig:Domains}\textbf{b}. Using the same fitting procedure as for the experimental data in Fig.\,\ref{fig:PhaseTransition}\textbf{b}, we find that the first critical QZE predicted by the GP approach is $q_1^{\textrm{GP}}\approx h \times 0.36~$Hz. Therefore the discrepancy between the measured and predicted first critical QZE is not resolved by upgrading the theory from LDA to GP.

A second critical QZE $q_2^{\textrm{GP}}\approx h \times 2~$Hz where $m_F=-1$ disappears can also be identified in the GP calculation. This is consistent with the experimental results, although we find experimentally that the population of the $m_F=-1$ component decreases smoothly with $q$ and does not completely vanishes at high $q$. This prevents us to clearly identify a critical value $q_{2,{\rm exp}}$ analogous to $q_2^{\textrm{GP}}$.

 \subsection{Role of the Thermal Components}

The discrepancy between the measured $q_1$ and the $T=0$ prediction, as well as the difficulty in identifying $q_2$ in experiments, can be understood qualitatively by considering the role of a finite temperature of the sample. To compare the experimental results with the prediction of the spin-1 GP theory in more detail and discuss the role of a thermal component, we define an effective size for each Zeeman component as the root-mean-square (rms) radius restricted to the condensate region $[-L,L]$, 
\begin{align}
R_{m_F} &  = \frac{1}{\mathcal{N}_{m_F}} \int_{-L}^L x^2 \rho_{1{\rm d},m_F}(x)dx,
\end{align}
where the half-length $L$ of the condensate is found by a parabolic fit to ${\rho_{1{\rm d}}(x)}$, and with a normalization factor $\mathcal{N}_{m_F}=\int_{-L}^L \rho_{m_F,1{\rm d}}(x)dx$. We show in Fig.\,\ref{fig:Fits}\textbf{a} the size $R_{m_F}$ computed from the measured profiles and from the calculated ones. The size of $m_F=+1$ increases only slightly with $q$, and stays close to the $T=0$ GP prediction for all values of $q$. In contrast, both $R_{m_F=0}$ and $R_{m_F=-1}$ differ substantially from the $T=0$ predictions. Focusing on $m_F=0$, the rms radius starts from a large value at low $q$, then decreases above $q_{1,\textrm{exp}}$ before settling to an asymptotic value above $q\gtrsim h\times2\,$Hz. The agreeement between experiments and $T=0$ theory improves with increasing $q$. 

The differences between experiment and theory can be explained qualitatively by thermal excitations. Low-energy excitations of homogeneous spin-1 BECs have been studied using the Bogoliubov approach \cite{kawaguchi2012a,stamperkurn2013a,phuc2011a}. In general, one expects for $q \neq 0$ that the Bogoliubov spectrum consists of three modes. For low values of $q \ll q_1$, where the (quasi-)condensate occupies the $m_F=\pm1$ states, one spin mode essentially reduces to excitations of atoms in the $m_F=0$ state with a gap $E_g \geq q$\,\cite{kawaguchi2012a}. In a Hartree-Fock picture appropriate for $k_BT \gg g_s \rho$, the effective potential seen by the uncondensed $m_F=0$ atoms is almost flat (up to small terms $\propto g_s$): The mean-field from the condensate in $m_F=\pm1$ cancels almost exactly the trapping potential \cite{scherer2010a,olf2015a}. Uncondensed excitations in $m_F=\pm 1$ experience a different mean-field potential that expels them from the trap center. As a result one expects that below $q_1$ the thermal component occupies mostly the $m_F=0$ Zeeman state. In Fig.\,\ref{fig:Fits}\textbf{b}, we show a magnified view of the linear density profiles for $q<q_{1,\textrm{exp}}$. A subtantial population is present in $m_F=0$ (in contrast to the $T=0$ prediction) and shows a ``flat-top'' profile within the volume where the $m_F=\pm1$ condensate is present, in agreement with the Hartree-Fock description. For a flat density confined within the condensate region $[-L,L]$, the rms radius is $\approx \sqrt{1/3}L \approx 62\,\mu$m, in good agreement with the measured $R_{m_F=0}\approx 58 \,\mu$m for low $q$. 

This discussion, although qualitative, explains the increase of the observed critical field from the $T=0$ value. For $q\gtrsim q_1$, the small domain expected at $T=0$ does not actually form but rather dissolve inside the existing $m_F=0$ thermal component. The suppression of phase separation at finite temperatures has been noted in a theoretical study of a two-component gas~\cite{roy2015a}, and is also consistent with our previous experimental work on three-dimensional spin-1 gases~\cite{frapolli2017a}. We are not aware of theoretical studies of antiferromagnetic spin-1 gases in 1d that can explain our observations quantitatively. Our experiments could be modelled using, \textit{e.g.}, classical field methods (reviewed \textit{e.g.} in \cite{blakie2008a}) and perhaps used to benchmark such methods. To ease such comparison, we have measured the temperature of the thermal component by fitting the equation of state obtained from the ``wings'' of the linear profiles \cite{ho2009a} to a Hartree-Fock model of our quasi-1d gas\,\cite{trebbia2006a}. Here the ``wings'' correspond to the non-degenerate region of the cloud where the one-dimensional phase space density $\rho_{\textrm{1d}}\lambda_T \leq 1$, with $\lambda_T=\sqrt{2\pi\hbar^2/(m_{\textrm{Na}}k_B T)}$ the thermal De Broglie wavelength and $k_B$ the Boltzmann constant. We find $T\approx 30-40\,$nK without any obvious dependence on $q$. Note that the measured temperature is substantially higher than the spin-dependent energies $\eta, q$ explored in this work.

\subsection{Long time relaxation of out-of-equilibrium spin textures}

\begin{figure*}
\includegraphics[width=1\textwidth]{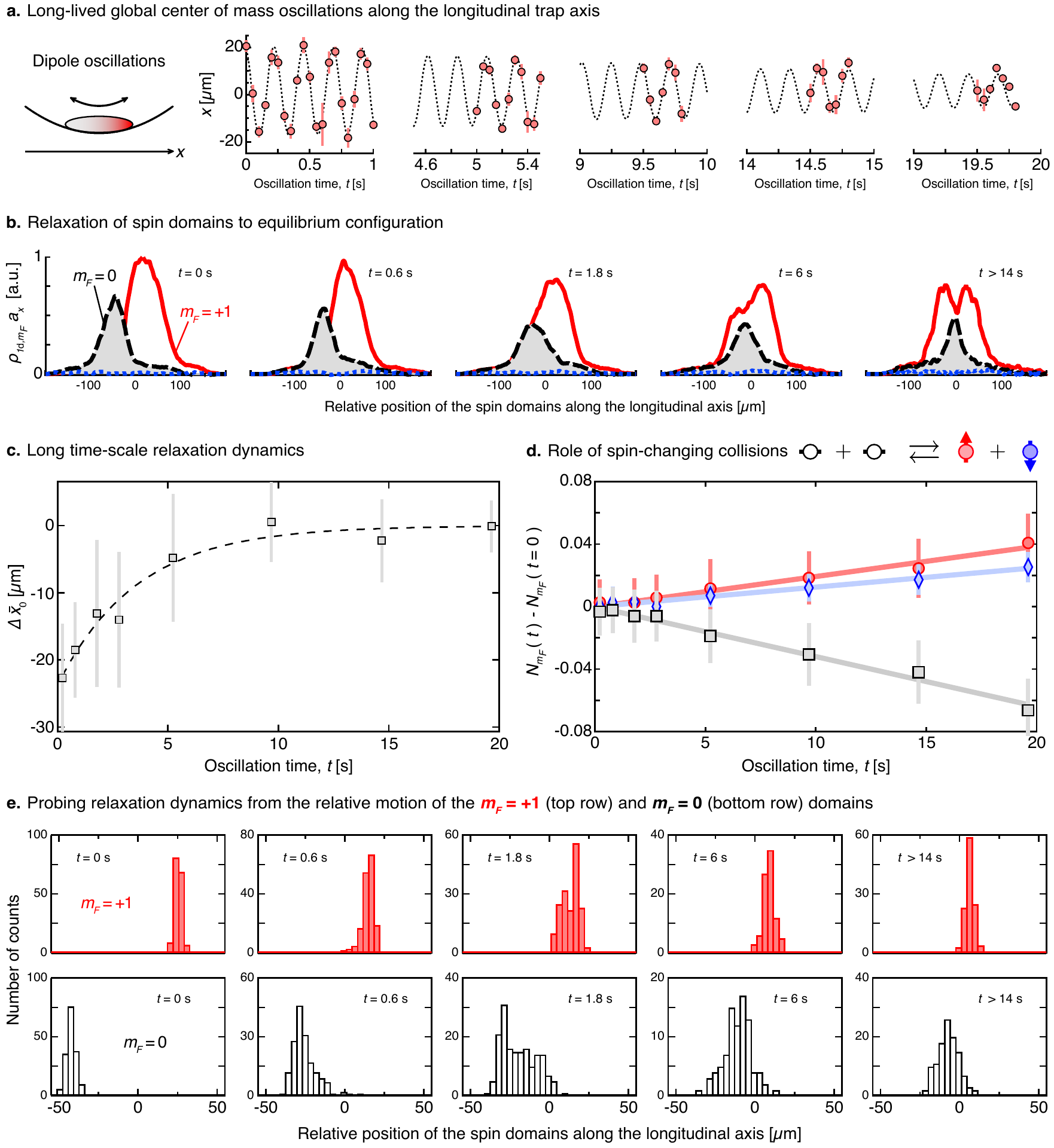}%
\vspace{-0pt}
\caption{{\bf Relaxation of spin domains to their equilibrium configuration.} {\bf a.} Long-lived center-of-mass oscillations of a partially magnetized gas with $M_{\|} /N\approx 0.66(2)$. {\bf b.} Relaxation to the final equilibrium configuration. An initially asymmetric equilibrium state prepared with $b'\neq0$ relaxes to a symmetric equilibrium state expected for $b'=0$ when the gradient $b'$ is suddently turned off. The relaxation is much slower than the axial trap period and spans several seconds. {\bf c.} Evolution of the center-of-mass $\Delta\overline{x}_0$ of the $m_F=0$ component relative to the center-of-mass of the whole cloud. An exponential fit to the data (dashed line) returns a $1/e$ relaxation time of \textbf{$\tau\approx 3.7(1.7)\,$s}.  {\bf d.} Relative change of the spin populations, $N_{m_F}(t)-N_{m_F}(0)$ (squares: $m_F=0$, circles: $m_F=+1$, diamonds: $m_F=-1$), indicating spin-changing dynamics is involved in the relaxation process. The straight lines are a guide to the eye. {\bf e.} Histograms of the relative displacement $\Delta\overline{x}_0$. Each histogram contains at least 170 measurements. For these measurements, the trap frequencies are $(\omega_x,\omega_\perp)=2\pi\times(4.3,385)\,$Hz and the atom number is $N\approx 9400$. Error bars in {\bf c,d} indicate the empirical standard deviations of the data.}
\label{fig:Dynamics}
\end{figure*}

Having characterized the equilibrium properties of a spin-1 antiferromagnetic gas, we now turn to non-equilibrium behavior. 
We investigate how an initial, highly non-equilibribrium configuration relaxes to a final equilibrium configuration. The experiment is performed at a uniform bias field $B=600\,$mG ($q/h \approx 100\,$Hz), well above 
$q_2^{\textrm{GP}}$. We prepare the system at a magnetization $M_{\|} \approx 0.66(2)\,N$ using the same procedure as before, except for an applied magnetic potential $V_{\mathrm{mag}}=g_F m_F \mu_B  b' x$ along $x$ controlled by an applied magnetic gradient $b'$ ($\mu_B$ is the Bohr magneton and $g_F=-1/2$ the Land\'e $g-$ factor). Using $b'=24\,$mG/cm, the net effect of the combined action of the magnetic force and of spin-dependent interactions is to pull the $m_F=+1$ Zeeman component to the right side of the cloud while pushing the $m_F=0$ component to the left one. Atoms in $m_F=-1$ are purely thermal and barely discernible in this regime. 

We remove the applied magnetic force at $t=0$, leaving the spin-1 gas in a purely optical potential independent of the Zeeman state but also in a highly non-equilibrium configuration. The first consequence is an excitation of the center-of-mass (c.o.m.) motion of the cloud that persists up to 20\,s, the longest time we explored (see Fig.\,\ref{fig:Dynamics}{\bf a}). This motion is common-mode to the $m_F=0$ and $+1$ components and occurs at the expected dipole mode frequency $\omega_x$. In contrast, the relative positions of the two Zeeman components do not display any detectable oscillation and evolve on a much longer time scale than the axial period, as pictured in Fig.\,\ref{fig:Dynamics}{\bf b}. To quantify the relaxation we introduce the c.o.m. displacements
\begin{align}
\Delta\bar{x}_{m_F} & = \frac{1}{N_{m_F}} \int\, (x-\bar{x})\rho_{1{\rm d},m_F}(x)dx, 
\end{align}
of the $m_F=+1$ and $m_F=0$ components from the center of mass $\bar{x} = (1/N) \int\, x\rho_{1{\rm d}}(x)dx$ of the whole cloud. Here $N_{m_F}=\int\,\rho_{1{\rm d},m_F}(x)dx$ is the total population of the $m_F$ component. We report in Fig.\,\ref{fig:Dynamics}{\bf c} the relative displacement of $m_F=0$, which remains mostly constant for several periods of the c.o.m. oscillations before decaying to zero within a timescale of $\sim 4\,$s. 

The profiles shown in Fig.\,\ref{fig:Dynamics}{\bf b} indicate that this relaxation occurs progressively, with the $m_F=0$ component penetrating slowly into the $m_F=+1$ majority component. This behavior could be surprising for a truly immiscible binary mixture, where the repulsion between the species acts as an effective barrier preventing relaxation. Fig.\,\ref{fig:Dynamics}{\bf d} shows that the $m_F=-1$ component, altough weakly populated, still plays a role in the relaxation process. The relative populations of the Zeeman states evolve in time on the same scale as the relaxation takes place, with a decrease in the population of $m_F=0$ and a roughly equal increase in the populations of $m_F=\pm 1$. This indicates that spin-changing collisions of the form $2\times (m_F=0) \longrightarrow (m_F=+1) + (m_F=-1)$ are involved in the mechanism enabling $m_F=0$ atoms to cross the effective energy barrier due to spin-dependent interactions. The process is most likely dominated by excitations (presumably thermal) residing initially in the inferface between the $m_F=0$ and $m_F=+1$ regions, and seeding the long-time dynamics\,\cite{stamperkurn1999b}. 

Fig.\,\ref{fig:Dynamics}\textbf{e} displays histograms of the c.o.m. of the $m_F=0,+1$ components as a function of the relaxation time. We observe a gradual change over time from a distributions peaked near the cloud edges to distributions peaked near the cloud center. 
The distribution of $\overline{x}_0$ appears smooth and single-peaked at all times. These observations rule out a scenario where relaxation is explained by a macroscopic quantum tunneling of the $m_F=0$ component. In that case, we expect at intermediate times that the $m_F=0$ component is in a superposition of two domains, one localized on the left side of the $m_F=+1$ cloud and one localized near its center. This would lead to a bimodal spatial distribution for which we find no evidence. 


\section{Discussion}

In summary we have investigated a spin $F=1$ Bose gas with antiferromagnetic interactions in a quasi-1D configuration. We measured the magnetic phase diagram in a uniform bias magnetic field. The applied bias field favors the appearance of $m_F=0$ atoms through the associated QZE and competes with spin-dependent interactions in a partially magnetized sample, where the low-field configuration is a mixed phase of the $m_F=\pm 1$ components. We experimentally found the critical value $q_{1,\textrm{exp}}$ where the $m_F=0$ domain appears.

We found that the $T=0$ mean-field theory of spin-1 Bose gases describes qualitatively well our observations. However there exist discrepancies between the predicted and measured values of the critical fields. The finite temperature of our samples, although very low, could explain these discrepancies. Indeed, energy scales in spinor gases are naturally low in comparison to the ``natural'' scale set by the chemical potential of the BEC. Therefore we expect that thermal fluctuations are able to suppress the formation of  spin domains near the transition where different spin configurations are close in energy. The quasi-1D nature of our experimental system may further enhance thermal effects.

Finally, we studied the non-equilibrium dynamics and relaxation of spin domains in the phase-separated, high-$q$ regime. In contrast to the miscible regime~\cite{Kim2017a,Bienaime2016a}, we observe no spin-dipole oscillations in the phase-separated regime. Instead we find that spin dynamics is frozen on short time scales on the order of the trap period, and undergoes slow relaxation towards an equilibrium configuration on long times scales of several tens of axial trap periods (about 10\,s). We found evidence that relaxation takes place through spin-changing collisions, enabling atoms from immiscible Zeeman components to ``pass through'' the effective barrier created by mean-field interactions with the other component. Our results could be explained by a thermally-assisted process where a scarcely populated, but not empty thermal component in $m_F=-1$ seeds the relaxation dynamics. We found no clear evidence of macroscopic quantum tunneling.



\section{Methods}

{\noindent {\bf Optical dipole trap.}} Our experiments start with a spinor gas of ultracold $^{23}$Na atoms with a fixed total magnetization $M_{\|}$ and immersed in a uniform magnetic field $\bm{ B}$. 
The spinor gas is held in a crossed dipole trap created at the intersection of two Gaussian beams propagating along the $x$ and $z$ axes. After achieving a degenerate Bose gas using standard evaporative cooling, we transfer the cloud in the 1d trap by adiabatically turning off one of the dipole beams in 5\,s (see Supplementary Material --SM-- for more details).

{\noindent {\bf Stern-Gerlach imaging.}} We measure the density profiles of each Zeeman component by removing suddenly the trapping potential and letting the cloud expand for a time-of-flight (t.o.f.) of $t=8\,$ms in a magnetic field gradient (applied only during the t.o.f.). Owing to the large trap anisotropy, the expansion is essentially in the radial direction (At $T=0$, the condensate expands along its weak axis by a factor $\approx 10^{-4}$ \cite{castin1996a}). The domain walls of width $\zeta$ associated with spin domains carry kinetic energy, and are therefore expected to expand at a speed $\sim \hbar /(m_{\textrm{Na}} \zeta) $ during the short t.o.f.~\cite{stamperkurn1999b}. However, in our experiments we have $\hbar t /(m_{\textrm{Na}} \zeta) \approx  1-2\,\mu$m$\ll \zeta$, so that we can safely neglect this expansion. 

{\noindent {\bf Magnetic field generation.}} We generate uniform magnetic fields using three pairs of bias coils aligned along the $x\pm y$ and $z-$directions. We calibrate the magnetic fields using radio-frequency spectroscopy, with a typical resolution of $\sim1\,$mG. We observe magnetic field fluctuations with $\delta B\sim 3\,$mG root-mean-square (r.m.s.) amplitude and with a typical time scale of several tens of seconds. These fluctuations, coming from a nearby subway line, are along the vertical $z$ axis, orthogonal to the applied bias field ${\bf B}$ that lies in the $x-y$ plane. The impact of magnetic field fluctuations is minimized by working with applied fields $B \geq 30\,$mG. The resulting r.m.s. uncertainty on $q$ is then below $\delta q \sim (\delta B/B)^2 \sim 1\,$\%.  
%

{\noindent {\bf Longitudinal magnetic force cancellation.}} Our experiments are performed after carefully cancelling stray magnetic field gradients (thereby cancelling magnetic forces) along the weak axis of the trap. Stray gradients have at least two origins: ({\it i}) the residual ambient gradients (arising from inhomogeneously magnetized elements around the experiment, power supplies, etc ...) and ({\it ii}) the imperfections of the bias coils that produce slightly inhomogeneous fields. 
We cancel the residual magnetic force along $x$ by two methods, either by applying a compensation gradient along the weak axis of the trap (more appropriate at low bias fields where effect ({\it i}) dominates), or by choosing the direction of the applied field (more appropriate at large bias fields where effect ({\it ii}) dominates) [see SM for more details]. We are able to cancel longitudinal magnetic gradients  to better than a few $100\,\mu$G/cm along the weak trapping direction $x$. Residual magnetic forces along the $y-$ and $z-$directions are negligible due to the larger confinement 

%
%

{\noindent {\bf Spin-1 Gross-Pitaevskii equations.} 
In the 1d limit, the complete BEC wavefunction can be written as ${\bf \Psi}=\phi_\perp(y,z){\bf \zeta}(x)$ where $\phi_\perp(y,z)$ denotes the transverse harmonic oscillator ground state. The one-dimensional spin-1 Gross-Pitaevskii equation can be written as a set of three equations for each Zeeman component, of the form 
\begin{align}
i \hbar \frac{\partial \zeta_{+1}}{\partial t} & =  \left[\mathcal{L} + g_{s}^{\textrm{1d}} ( \rho_{\textrm{1d},0}+ m_{\|} )\right]\zeta_{+1} +g_{s}^{\textrm{1d}} \zeta_{0}^2 \zeta_{-1}^\ast, \label{GPP1}\\
\nonumber
i \hbar \frac{\partial \zeta_{0}}{\partial t} & = \left[ \mathcal{L} + g_{s}^{\textrm{1d}} ( \rho_{\textrm{1d}}-\rho_{\textrm{1d},0}) \right] \zeta_{0} + 2g_{s}^{\textrm{1d}} \zeta_{0}^\ast \zeta_{-1}\zeta_{+1},\\
\nonumber
i \hbar \frac{\partial \zeta_{-1}}{\partial t} & =  \left[\mathcal{L} + g_{s}^{\textrm{1d}} ( \rho_{\textrm{1d},0} - m_{\|}) \right]\zeta_{-1} +g_{s}^{\textrm{1d}} \zeta_{0}^2 \zeta_{+1}^\ast,  
\end{align}
with $ \mathcal{L}=\hat{h}+\overline{g}^{\textrm{1d}} \rho_{\textrm{1d}}$ the spin-independent GP operator and $m_{\|}=\rho_{\textrm{1d},+1}-\rho_{\textrm{1d},-1}$ the density of longitudinal magnetization. 

We propagate Eqs.\,(\ref{GPP1}) in imaginary time to obtain the lowest energy state using a split-step method. The evolution due to the kinetic energy, local spin-independent and local spin-dependent terms are calculated separately by exponentiating the corresponding operator. This can be done analytically, either in the momentum or position basis. Then the total evolution at each time step is approximated by multiplying all three evolution operators neglecting their commutation properties (first-order Trotter expansion). We have studied the influence of the time step carefully to make sure the higher-order terms are indeed negligible. 

We use harmonic oscillator units where time is rescaled by $\omega_x^{-1}$, energy by $\hbar\omega_x$, and lengths by $a_x=\sqrt{\hbar/(m_{\textrm{Na}}\omega_x)}$. For the data shown in this paper, we typically use a grid containing $N=64$ points with grid spacing $\Delta x =15/32$, an imaginary time step $\delta t =2\cdot 10^{-5}$ and we compute the imaginary time evolution up to $T=10^3$. We use dimensionless coupling constants $N\overline{g}^{\textrm{1d}} = N \overline{g}/( 2\pi \hbar \omega_x a_\perp^2 a_x) \approx 378.9$ and $g_s^{\textrm{1d}}/\overline{g}^{\textrm{1d}} = 0.0357 $.

\section{Data availability}
\noindent The data that support the findings of this study are available from the corresponding author upon request.

\section{Acknowledgements}
\noindent We acknowledge stimulating discussions with Emilia Witkowska, Sandro Stringari and Gabriele Ferrari. This work has been supported by ERC (Synergy Grant UQUAM). K.\,J.\,G. acknowledges funding from the European Union's Horizon 2020 Research and Innovation Programme under the Marie Sklodowska-Curie Grant Agreement No. 701894. LKB is a member of the SIRTEQ network of R\'egion Ile-de-France.

\section{Author Information}

\subsection{K. Jim{\'e}nez-Garc{\'i}a \& A. Invernizzi}
These authors contributed equally to this work.

\subsection{Affiliations} 
\noindent {\it Laboratoire Kastler Brossel, Coll{\`e}ge de France, CNRS, ENS-PSL Research University, Sorbonne Universit{\'e}, 11 Place Marcelin Berthelot, 75005 Paris, France}\\
K. Jim{\'e}nez-Garc{\'i}a, A. Invernizzi, B. Evrard, C. Frapolli, J. Dalibard \& F. Gerbier\\

\noindent {\it Centro de Investigaci{\'o}n y Estudios Avanzados del Instituto Polit{\'e}cnico Nacional - Unidad Quer{\'e}taro, 76230 Quer{\'e}taro, M{\'e}xico}\\
K. Jim{\'e}nez-Garc{\'i}a (current address)\\

\noindent {\it Safran, 2 Boulevard du G{\'e}n{\'e}ral Martial Valin, 75724 Paris, France}\\
A. Invernizzi, C. Frapolli (current address)

\subsection{Contributions} 
\noindent K.J.-G. and A.I configured the existing experimental apparatus to perform 1D experiments. K.J.-G., A.I and B.E. gathered the data with assistance from C.F. K.J.-G. and A.I carried out the analysis of the data. K.J.-G., A.I, J.D. and F.G performed numerical and analytical calculations. All authors contributed to discussions and preparation of the manuscript. This work was supervised by J.D. and F.G.

\subsection{Competing interests}
\noindent The authors declare no competing financial interests.

\subsection{Corresponding author}
\noindent Correspondence to Fabrice Gerbier (fabrice.gerbier@lkb.ens.fr).



\section{References}

\bibliography{Spin1D_bibliography_extended}

\end{document}